\documentclass[12pt]{iopart}

\usepackage{graphicx}

\begin{document}

\title[Analytic expression for pull-or-jerk experiment]{Analytic expression for pull-or-jerk experiment}
\author{Hiroyuki Shima}

\address{
Department of Environmental Sciences, University of Yamanashi, 4-4-37, Takeda, Kofu, Yamanashi 400-8510, Japan}
\ead{hshima@yamanashi.ac.jp}
\vspace{10pt}
\begin{indented}
\item[]\today
\end{indented}

\begin{abstract}
This work focuses on a theoretical analysis of a well-known inertia demonstration, which uses a weight suspended by a string with an extra string that hangs below the weight. The point in question is which string, the upper or lower, will break when pulling down on the bottom edge of the lower string at a constant pulling speed. An analytic solution for the equation of motion allows us to identify the critical value of the pulling speed, beyond which the string breaking varies from one to the other. The analysis also provides us a phase diagram that illustrates the interplay between the pulling speed and the string's elasticity in the pull-or-jerk experiment.
\end{abstract}

%

\submitto{\EJP}
%
%
%

\section{Introduction}

Suppose that a large and heavy object (weight) is suspended from a ceiling by a long light string, and a similar string is suspended from the bottom of the weight. If you jerk the lower string downward, the lower string breaks, but the upper string survives, and the weight remains suspended from the ceiling. In contrast, if you slowly pull the lower string downward, only the upper string breaks, and the weight falls off. These two contrasting observations, in other words a ``pull-or-jerk" experiment, originate from the inertia of the suspended weight; the inertia of the weight prevents the upper string from breaking when the lower string is jerked. In fact, this experiment has been repeatedly used in course of elementary physics as a concise demonstration of inertia.

\vspace*{12pt}

The underlying mechanism of the pull-or-jerk experiment is described below in a qualitative manner. A quick jerk causes a rapid increase in the tension of the lower string. Nevertheless, the inertia of the weight prevents an abrupt acceleration of the weight, which
results in a further elongation in the lower string. Eventually, the tension in the lower string reaches a maximum value, which results in
the lower string breaking when jerked quickly. In contrast, a slow pull will break only the upper string because of the non-equivalence in the tensions between the two strings. The tension in the upper string is a sum of the restoring force plus the gravitational force on the weight, while the tension in the lower string is only equals the restoring force. Therefore, the upper string will reach its maximum tension first when the lower string is pulled slowly.

\vspace*{12pt}

Despite the qualitative, stereotypical explanation given above, the story does not end there. Using stretchable strings make what we observe richer. As such, no matter how slowly or quickly we pull the lower string, the elastic properties of the two strings will cause an oscillatory motion of the weight along the vertical direction. The oscillation then induces a time modulation in the tensions of the two strings. As a consequence, the problem is beyond a two-alternative choice, i.e., either a jerk or slow pull. Depending on the pulling speed and elastic properties of the strings, how each string breaks may vary. Thus, we are required to analytically solve the equation of motion and derive the time-variation in the tensions, which will provide a good gedanken-experiment and facilitate a student's understanding of mechanics. Nevertheless, there have only been a few attempts at solving the problem in a quantitative manner \cite{LeCorbeiller1945,Karioris1978,Heald1996,Caplan2004}.

\vspace*{12pt}

This contribution provides an analytic description of the pull-or-jerk experiment by solving the equation of motion of the system. The analysis is based on the assumption that the bottom edge of the lower string moves linearly downward with time; this constant-pulling-speed condition is different from the existing studies, in which the pulling force acting on the bottom of the lower string varies
linearly \cite{LeCorbeiller1945,Heald1996,Caplan2004}
or sinusoidally \cite{Karioris1978}
with time. Special attention is paid to a closed-form solution for the critical pulling speed that causes the string to break from the upper to lower (or vice versa). Physical interpretation of key parameters that determine the critical pulling speed is also clarified.

\begin{figure}[ttt]
\centerline{\includegraphics[width=0.5\textwidth]{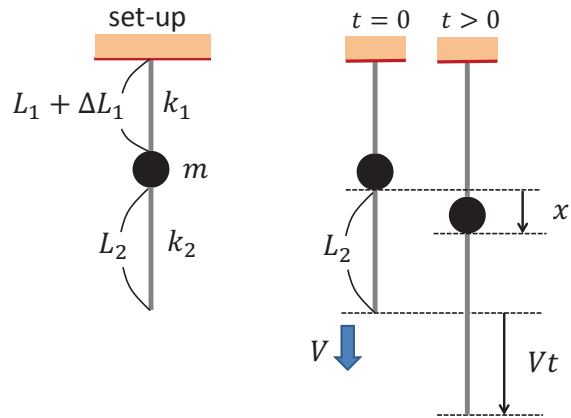}}
\caption{Theoretical model consisting of two elastic strings (with original lengths $L_1$ and $L_2$)
and a weight (with mass $m$).
The pulling speed $V$ (constant in time) and the displacement of the weight $x$
are positive in downward direction.}
\label{fig01}
\end{figure}

\section{Model}

Figure \ref{fig01} shows a schematic diagram of the theoretical model considered herein. A weight
with mass $m$ is suspended by a stretchable string with a spring constant
$k_1$ and original length $L_1$; another string with $k_2$ and
$L_2$ is attached to the weight and dangled below it. The
two strings can be assumed to be made of the same material without any loss of
generality. Although they have the same Young's modulus value, their
spring constants should be in proportion to their respective lengths. We
further assume that each string is massless and obeys Hooke's law up to the
failure at the maximum tension $f_{\rm max}$. The tension $f_i \;(i=1,2)$ in the $i$th string is expressed by
\begin{equation}
f_i=k_i \Delta L_i \;\; \mbox{(at $f_i < f_{\rm max}$)},
\end{equation}
where $\Delta L_i$ is an increment in the $i$th string's length.
Because the strings are massless, the tension in each string is the same throughout.

\vspace*{12pt}

A pull-or-jerk experiment is initiated by pulling down on the bottom of the
lower string with a constant speed $V$. Because of the elastic properties of
the two strings, the position of the weight $x$ oscillates with time
superposed on a linear ramp.
During the downward pulling ({\it i.e.}, at $t>0)$, the length of the lower string
is expressed by $L_2+Vt-x$, as seen from the right panel in Fig.~\ref{fig01}.
Hence, the tension that
arises in the lower string is given by
\begin{equation}
f_2 = k_2 \left[ \left( L_2+Vt-x \right)-L_2 \right]=k_2\left( Vt-x \right).
\end{equation}
The tension of the upper string is also dependent on $x$ and is expressed as
\begin{equation}
f_1 = k_1 x + mg.
\end{equation}
Ignoring damping effects for simplicity,
we obtain the equation of motion with respect to the weight as
\begin{equation}
m\frac{d^2 x}{dt^2}=-\left( k_1x+mg \right)+mg+k_2(Vt-x),
\end{equation}
which can be rewritten as
\begin{equation}
\frac{d^2 x}{dt^2}+\frac{k_1+k_2}{m}x=\frac{k_2 V}{m}t.
\label{eq_01}
\end{equation}
Imposing the initial condition of $x=0$ and $dx/dt=0$ at $t=0$, we obtain
the solution of Eq.~(\ref{eq_01}) as
\begin{equation}
x = -\frac{k_2 V}{m\omega^3}\sin \omega t + \frac{k_2 V}{m\omega^2}t,
\label{eq_02}
\end{equation}
with the definition of
\begin{equation}
\omega =\sqrt \frac{k_1+k_2}{m}.
\end{equation}

\vspace*{12pt}

The solution given by Eq.~(\ref{eq_02}) implies that the tensions $f_1$ and $f_2$ are in the forms of
\begin{eqnarray}
f_1 &=&  -\frac{k_1 k_2 V}{m\omega^3} \sin \omega t + \frac{k_1 k_2 V}{k_1+k_2}t + mg,\nonumber \\
f_2 &=&   \frac{k_2^2 V}{m\omega^3}\sin \omega t + \frac{k_1 k_2 V}{k_1+k_2}t.
\label{eq_03}
\end{eqnarray}
For convenience, we introduce a normalized ({\it i.e.,} dimensionless) pulling speed $v^*$, which was defined by
\begin{equation}
v^* = \frac{k_2 V}{m\omega g},
\end{equation}
in addition to the two parameters
\begin{equation}
P_1 = \frac{k_1 g}{\omega^2},\;\;
P_2 = \frac{k_2 g}{\omega^2}. \quad
\left(
\frac{P_i}{mg}=\frac{k_i}{k_1+k_2},\;\;
\frac{P_1+P_2}{mg}=1
\right)
\end{equation}
The physical meanings of $v^*,\, P_1, \, P_2$ will be discussed later.
Using this nomenclature, Eq.~(\ref{eq_03}) can be rewritten in a simple form as
\begin{eqnarray}
f_1 &=& v^* P_1 \left( \omega t - \sin \omega t \right) + mg, \nonumber \\
f_2 &=& v^* P_1 \cdot \omega t + v^* P_2 \cdot \sin \omega t,
\label{eq_04}
\end{eqnarray}
or normalized as
\begin{eqnarray}
\frac{f_1}{mg} &=& v^* \frac{P_1}{mg} \left( \omega t - \sin \omega t \right) + 1,\\
\frac{f_2}{mg} &=& v^* \frac{P_1}{mg} \left( \omega t + \frac{P_2}{P_1}\sin \omega t \right).
\end{eqnarray}

\vspace*{12pt}

These analytic formulations of $f_1$ and $f_2$ provide an important
result on the pulling speed $v^*$. Suppose that $f_1=f_2$ at
a specific time $t=t_{\rm c}$ that satisfies the relation
\begin{equation}
- v^* P_1 \sin \omega t_{\rm c} + mg = v^* P_2 \sin \omega t_{\rm c}.
\end{equation}
This implies that
\begin{equation}
\sin \omega t_{\rm c} = \frac{1}{v^*} \cdot \frac{mg}{P_1+P_2} = \frac{1}{v^*}.
\label{eq_05}
\end{equation}
Therefore, $v^*\ge 1$ is necessary for the existence of $t=t_{\rm c}$,
across which the magnitude relationship between $f_1$ and $f_2$ is
reversed. In other words, $v^* \ge 1$ is a necessary condition for the
occurrence of the transition from a ``slow-pull" to ``sharp-jerk" situation
(and vice versa).

\vspace*{12pt}

\begin{figure}[ttt]
\centerline{\includegraphics[width=0.8\textwidth]{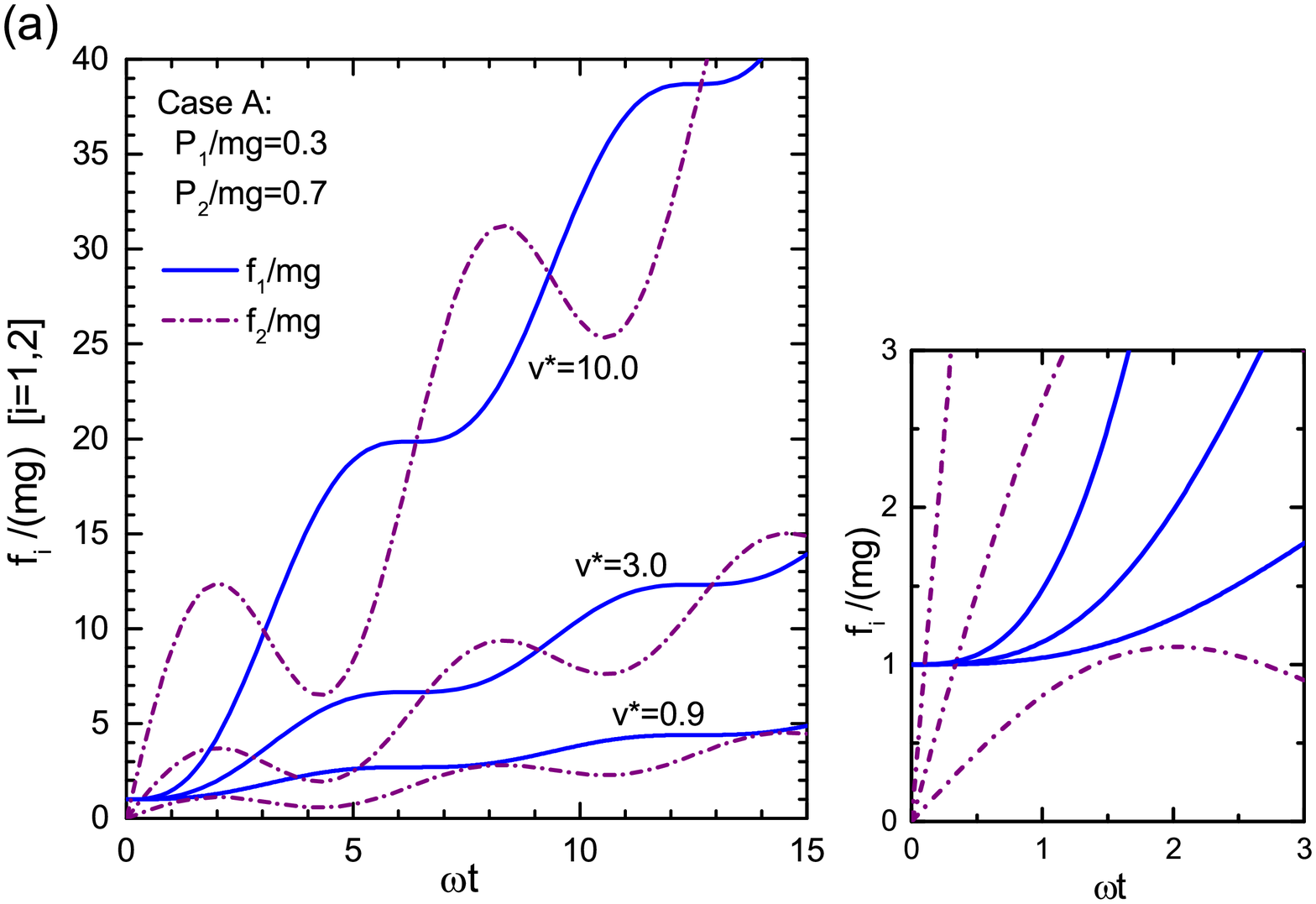}}
\centerline{\includegraphics[width=0.8\textwidth]{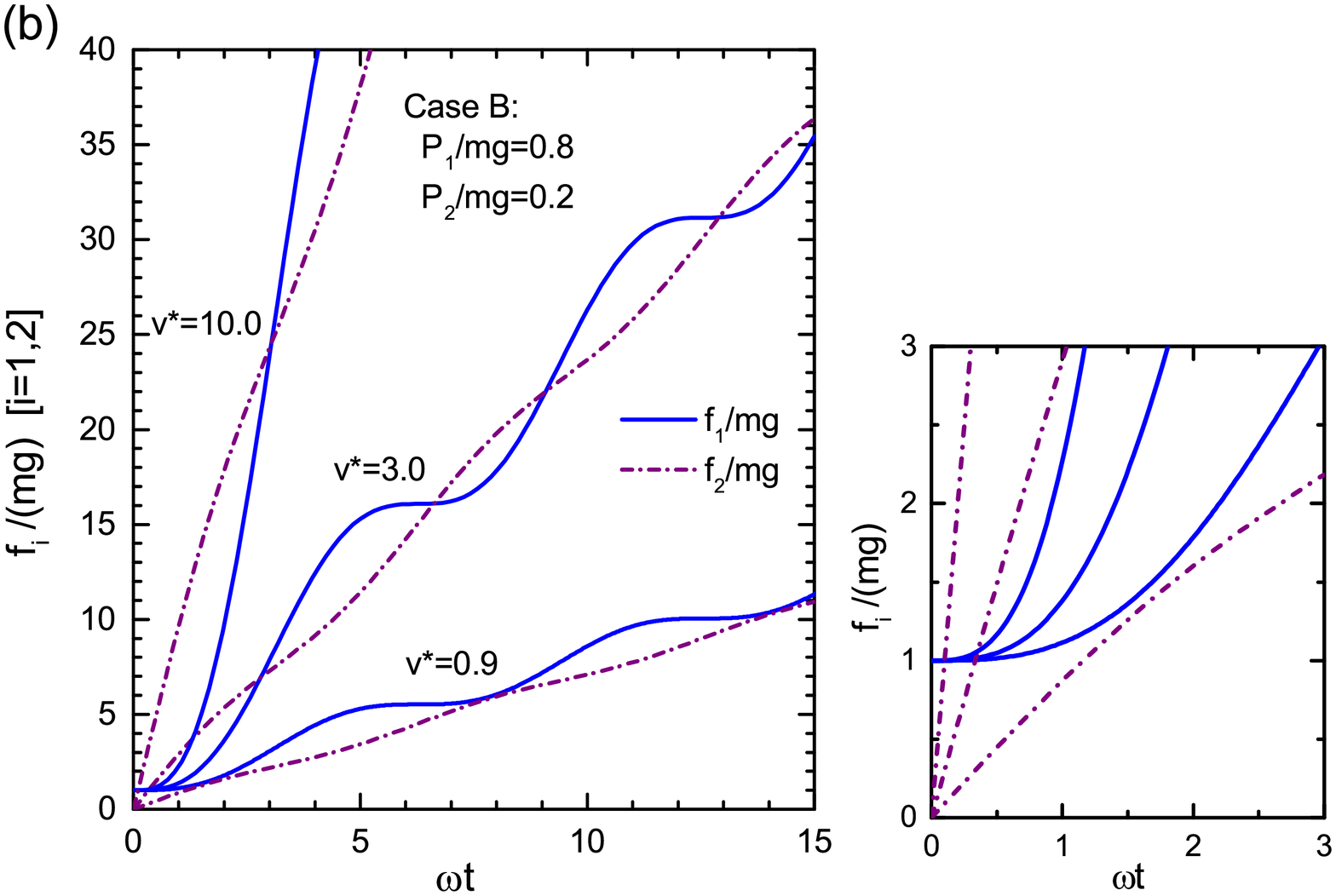}}
\caption{Time development of the tensions $f_1$ and $f_2$ in the upper and lower string, respectively.
Parameters of $v^*, P_1, P_2$ are varied as indicated in the plots.
Inset: Enlarged view for the small $t$ region. Absence of crossing points between the two curves
$f_1(t)$ and $f_2(t)$ for $v^*<1$ is clarified in the eye.}
\label{fig02}
\end{figure}

Figure \ref{fig02} shows the time evolution of the normalized tensions $f_1/mg$ and
$f_2/mg$ with respect to $\omega t$. The parameter $P_1/mg$ is fixed as 0.3 in Fig.~\ref{fig02}(a) and 0.8 in Fig.~\ref{fig02}(b); the relation of $P_2/mg=1-P_1/mg$
follows from the definition. When $v^*<1$, two oscillating curves of $f_1/mg$ and $f_2/mg$ have no crossing
point (see the inset of Fig.~\ref{fig02}), which is expected from Eq.~(\ref{eq_05}).
Because $f_1>f_2$ at any moment for the slow-pull case, the upper string must break regardless of
the $f_{\rm max}$ value. This scenario fails
if we speed up the pulling such that $v^*>1$. For the jerk case,
the two curves of $f_1/mg$ and $f_2/mg$ intersect periodically with
time, and $f_2$ can exceed $f_1$ at a certain moment.
As a consequence, the breaking string depends on the
value of $f_{\rm max}$ as well as $v^*$. These dependencies are the main issue that we address in the subsequent discussion.

\vspace*{12pt}

\section{Physical interpretation of parameters introduced}

\begin{figure}[ttt]
\centerline{\includegraphics[width=0.55\textwidth]{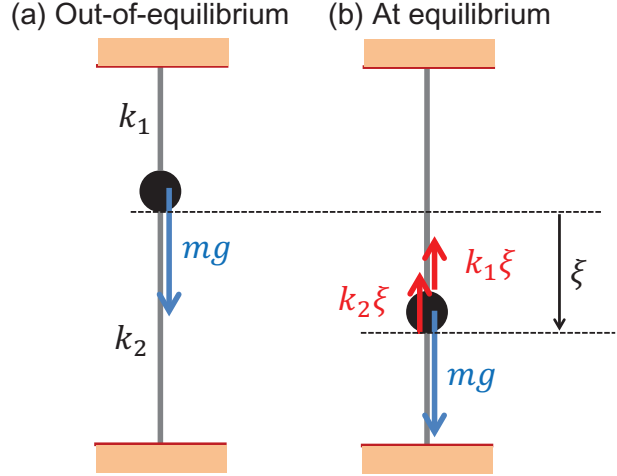}}
\caption{A weight vertically clamped by strings both from above and below.
(a) The system is out of equilibrium, {\it i.e.}, no force is exerted on the two strings.
At the next moment, the weight will fall due to the gravitational force $mg$.
(b) The system is in equilibrium, {\it i.e.}, the upper (or lower) string is elongated (contracted)
by $\xi$. The weight remains at rest because of the balanced forces.}
\label{fig03}
\end{figure}

Before proceeding with the argument, we discuss the
physical meanings of $v^*, P_1$, and $P_2$ that were introduced in Eq.~(\ref{eq_04}).
From their definitions, we can say that
\begin{equation}
v^* = \frac{k_2 V \cdot (1/\omega)}{mg},
\end{equation}
\begin{equation}
P_i=\frac{k_i}{k_1+k_2}mg. \;\; [i=1,2]
\end{equation}
Namely, the normalized velocity $v^*$ is the ratio between the
gravitational force ({\it i.e.,} $mg$) and the restoring force
({\it i.e.,} $k_2 V \cdot [1/\omega]$) of the lower string elongated by a
constant-rate $V$ during a time $1/\omega$. $P_1$ and $P_2$ are the
restoring forces of the upper and lower strings, respectively, under the
condition that the system is clamped vertically at both ends (see Fig.~\ref{fig03});
the downward displacement $\xi$ of the weight causes the upward forces
$k_1 \xi$ and $k_2 \xi$ to exert a force on the weight. At equilibrium,
the balanced forces are expressed as
\begin{equation}
mg = \left( k_1+k_2 \right)\xi.
\end{equation}
Thus, it follows that
\begin{equation}
k_i \xi = \frac{k_i}{k_1+k_2}mg = P_i, \;\; [i=1,2]
\end{equation}
which indicates that $P_i$ is the upward force $k_i \xi$ mentioned above.

\vspace*{12pt}

\section{Critical pulling speed}

We are ready to derive a closed-form expression for the critical pulling speed.
Equations (\ref{eq_04}) and (\ref{eq_05}) tell us that $f_2$ at $t=t_{\rm c}$ can be written as
\begin{equation}
f_2(t=t_{\rm c}) = P_2 + v^* P_1 \sin^{-1}\left( \frac{1}{v^*} \right).
\end{equation}
The two strings will break simultaneously at $t=t_{\rm c}$ if $f_2(t=t_{\rm c})$ is equal to $f_{\rm max}$.
Therefore, the pull-to-jerk phase boundary is expressed by a series of points $\{ v^*, f_{\rm max} \}$
in the $v^*$-$f_{\rm max}$ space, at which the following relation is satisfied
\begin{equation}
P_2 + v^* P_1 \sin^{-1} \left( \frac{1}{v^*} \right) = f_{\rm max},
\end{equation}
or equivalently,
\begin{equation}
\frac{1}{v^*} = \sin \left( \frac{1}{v^*} \cdot \frac{f_{\rm max}-P_2}{P_1} \right).
\label{eq_06}
\end{equation}
To numerically solve Eq.~(\ref{eq_06}) with respect to $v^*$,
we define
\begin{equation}
G\left( v^* \right)=\sin \left( \frac{1}{v^*}\cdot
\frac{f_{\rm max}-P_2}{P_1} \right)-\frac{1}{v^*},
\label{eq_07a}
\end{equation}
which has dimensionless parameters $f_{\rm max}/mg$ and $P_1/mg$;
note again that $P_2/mg=1-\left(P_1/mg \right)$. Roots of the equation $G\left(
v^* \right)=0$ vary in response to a change in $f_{\rm max}$.
Therefore, collecting all the roots with increasing $f_{\rm max}$ 
allows us to draw the phase diagram of the pull-or-jerk experiment in the
$v^*$-$f_{\rm max}$ space. In the actual calculation, the roots of $G\left( v^* \right)=0$
are detected using a conventional Newton-Raphson method.

\begin{figure}[ttt]
\centerline{\includegraphics[width=0.75\textwidth]{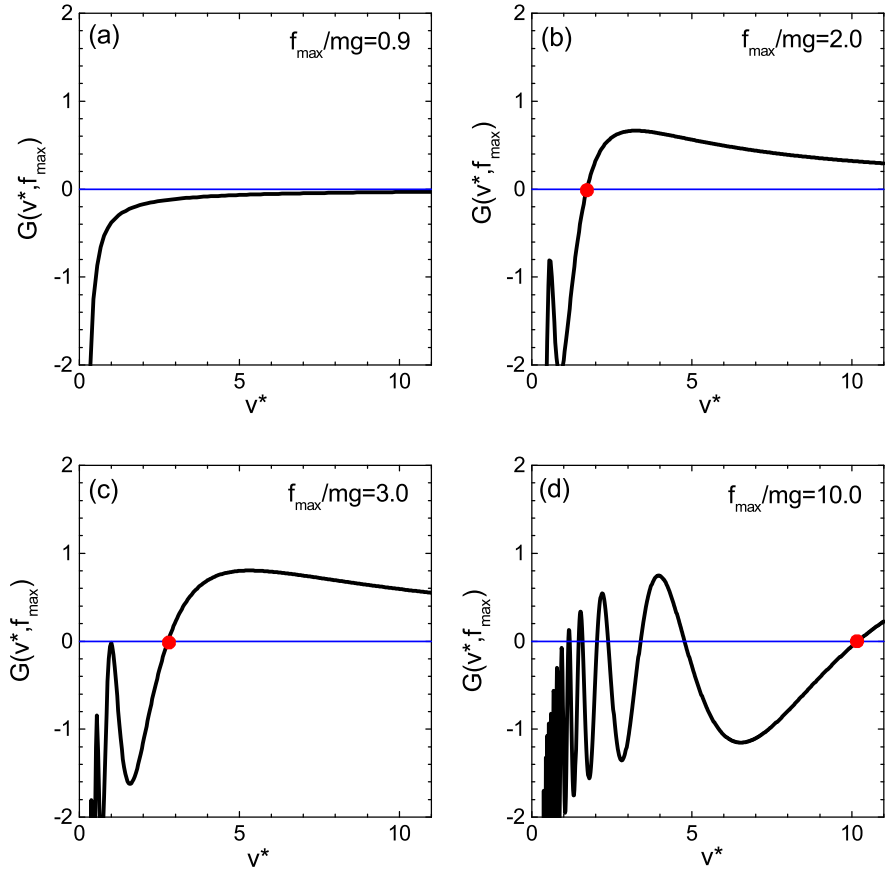}}
\caption{Evolution in the profile of $G(v^*)$ with increasing $f_{\rm max}$:
(a) $f_{\rm max}/mg=0.9$, (b) $f_{\rm max}/mg=2.0$, (c) $f_{\rm max}/mg=3.0$,
(d) $f_{\rm max}/mg=10.0$.
For all the four plots, $P_1/(mg)=0.3$ is fixed.
Solid circles (colored in red) trace the movement of the rightmost crossing point
for an increase of $f_{\rm max}$.}
\label{fig04}
\end{figure}

\vspace*{12pt}

Figure \ref{fig04} shows the profile of $G\left( v^* \right)$ for various conditions of $f_{\rm max}/(mg)$.
The same value of $P_1/(mg)$ as in Fig.~\ref{fig02}(a) is chosen for clarity.
For $f_{\rm max}/(mg) \le 1$, $G\left( v^* \right)<0$ for every $v^*$; therefore, no root of $G\left( v^* \right)=0$ can be found.
This absence of a root is obvious from both the mathematics and physics viewpoints.
From the mathematics viewpoint, $G\left( v^* \right)$ for $f_{\rm max}/(mg)\le 1$
is reduced to $G\left( v^* \right)=\sin \left( c/ v^* \right)- (1/v^*)$
with a constant $0<c\le 1$; thus, it is always negative \cite{comment}.
From the physics viewpoint, when $f_{\rm max}\le mg$,
the upper string does not sustain the weight from the outset, and the experiment itself fails;
hence, there is no room for the lower string to break.
Consequently, the pull-to-jerk transition
arises when $f_{\rm max}/(mg)>1$, which is signified by the presence of crossing points of the curve $G\left( v^* \right)$
with the $v^*$ axis in Figs.~\ref{fig04}(b)-\ref{fig04}(d).
At $f_{\rm max}/(mg)=2$ in Fig.~\ref{fig04}(b),
the crossing point at $v^*\sim 1.7$, which is marked by a solid circle, determines the critical
pulling speed; a quicker pull causes the lower string to break, and a slower
pull causes the upper string to break. Interestingly, the number of the
crossing points monotonically increase with the increase in $f_{\rm max}$.
In addition, they entirely shift to a larger $v^*$ with increasing
$f_{\rm max}$, as traced by solid circles that indicate the rightmost crossing point for each $f_{\rm max}/(mg)$.
This multiplicity in the roots of $G\left( v^*
\right)=0$ and their monotonic shift with a increase in  $f_{\rm max}$ are what we
have earlier commented in the Introduction as ``richness" of the pull-or-jerk experiment that are often overlooked in
elementary physics courses.

\vspace*{12pt}

\begin{figure}[ttt]
\centerline{\includegraphics[width=0.6\textwidth]{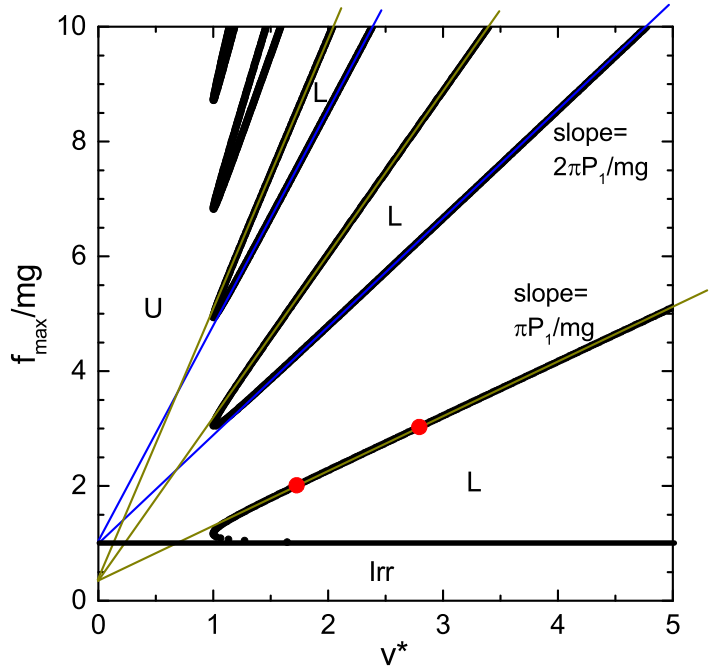}}
\caption{Phase diagram of the pull-or-jerk experiment under the constant-pulling-speed condition.
The symbols ``U" and ``L" indicate the phase domain at which the upper and lower string breaks, respectively.
The domain ``Irr" corresponds to an irrelevant situation that the string is too fragile to sustain the weight, and the experiment fails.
See text for thin slanted lines (colored in blue or yellow) that are superposed on the boundary curves.}
\label{fig05}
\end{figure}

\section{Phase diagram}

Figure \ref{fig05} shows a phase diagram describing the domains of the upper-string-break phase (designated by ``U")
and lower counterpart (``L") in the $v^*$-$f_{\rm max}$ space.
The undermost domain below the $f_{\rm max}/(mg)=1$ line, which is marked by ``Irr," corresponds to an irrelevant situation
in which the string is too fragile to sustain the weight, and the experiment fails.
We observe that multiple branches of the U-domain spread out from bottom left to top right in a radial manner,
and gaps between adjacent branches are occupied by L-domains.
All the plots along the boundary curves correspond to the roots of $G(v^*, f_{\rm max})=0$; the
two particular points highlighted in Fig.~\ref{fig05} by solid circles
(colored in red) are those depicted in Figs.~\ref{fig04}(b) and \ref{fig04}(c).
Each boundary curve asymptotically approaches a straight line for $v^*\gg 1$,
whereas in the vicinity of $v^*=1$ they are curvilinear.

\vspace*{12pt}

An interesting observation in Fig.~\ref{fig05} is the proportional relation in the slope of the asymptotic straight lines at $v^* \gg 1$.
To decribe the relation, 
we label the lines as $j=0,1,2,\cdots$ from the bottom, such that the $0$th (horizontal) and $1$st
(slanted) lines sandwich the lowermost L-domain immediately above the Irr-region.
Next, we denote the slope of the $j$th asymptotic straight line by $\gamma_j$.
It is then found that
\begin{equation}
\gamma_0 : \gamma_1 : \gamma_2: \gamma_3: \cdots
=
0:1:2:3:\cdots.
\end{equation}
This proportionality rule holds even when other values of $P_1/(mg)$ are considered.

\vspace*{12pt}

Further, the position of the intercept that we obtain
if extrapolating the asymptotic lines into the $f_{\rm max}$-axis is noteworthy.
Denoted by $\eta_j$, {\it i.e.}, the intercept of the $j$th line,
it appears that $\eta_j = 1$ for $j=0,2,4,\cdots$
and $\eta_j$ is equal to a constant less than unity for $j=1,3,5,\cdots$.
Namely, there are only two possibilities for the intercepts on the $f_{\rm max}$-axis
into which the asymptotic lines extrapolate.
This phenomenon is visualized in Fig.~\ref{fig05}
by superposing thin lines colored in blue (for $j=0,2,4,\cdots$)
and yellow (for $j=1,3,5,\cdots$).

\vspace*{12pt}

The two interesting features as to the slope $\gamma_j$ and intercept $\eta_j$
can be accounted for by considering the properties of $G(v^*)$
given by Eq.~(\ref{eq_07a}).
Suppose that $v^*$ and $f_{\rm max}$ satisfy either of the two relations below:
\begin{equation}
\frac{f_{\rm max}-P_2}{P_1} = 2j \pi v^* + 1,
\label{eq_07}
\end{equation}
or
\begin{equation}
\frac{f_{\rm max}-P_2}{P_1} = (2j+1) \pi v^* - 1.
\label{eq_08}
\end{equation}
In the former case, we have
\begin{eqnarray}
G(v^*)
&=& \sin \left[ \frac{1}{v^*} \cdot (2j \pi v^* + 1) \right] - \frac{1}{v^*}
= \sin \left( \frac{1}{v^*} + 2j \pi \right) - \frac{1}{v^*} \nonumber \\
&=& \sin \left( \frac{1}{v^*} \right) - \frac{1}{v^*},
\end{eqnarray}
which converges to zero \cite{sinc} for $v^* \gg 1$.
Similarly, in the latter case, we have
\begin{eqnarray}
G(v^*)
&=& \sin \left\{ \frac{1}{v^*} \cdot \left[ (2j+1) \pi v^* - 1\right] \right\} - \frac{1}{v^*} \nonumber \\
&=& \sin \left[ (2j+1) \pi - \frac{1}{v^*} \right] - \frac{1}{v^*}
= \sin \left( \frac{1}{v^*} \right) - \frac{1}{v^*},
\end{eqnarray}
which again converges to zero for $v^* \gg 1$.
Namely, Eqs.~(\ref{eq_07}) and (\ref{eq_08}) represent the asymptotic lines depicted in Fig.~\ref{fig05},
the points of which satisfy $G(v^*, f_{\rm max}) \simeq 0$ at $v^*\gg 1$.
We next note that Eqs.~(\ref{eq_07}) and (\ref{eq_08}) can be rewritten as
\begin{equation}
\frac{f_{\rm max}}{mg} = 2j \frac{P_1}{mg} \pi v^* + 1
\label{eq_09}
\end{equation}
and
\begin{equation}
\frac{f_{\rm max}}{mg} = (2j+1) \frac{P_1}{mg} \pi v^* + 1 - 2 \frac{P_1}{mg},
\label{eq_10}
\end{equation}
respectively.
Therefore, the slope $\gamma_j$ and intercept $\eta_j$ for the $j$th asymptotic lines
can be written as
\begin{equation}
\gamma_j = j \frac{P_1}{mg} \pi \;\; \mbox{for $j=0,1,2,\cdots$}
\end{equation}
and
\begin{equation}
\eta_j = \left\{
\begin{array}{ll}
1 & \mbox{for $j=0,2,4,\cdots$},\\
1-2\displaystyle{\frac{P_1}{mg}} & \mbox{for $j=1,3,5,\cdots$}.
\end{array}
\right.
\end{equation}
These results explain the proportionality relation in $\{\gamma_j\}$,
and the alternative choices in $\eta_j$ that were demonstrated.

\vspace*{12pt}

\section{Experimental feasibility}

The following discussion helps us to check the feasibility of a classroom experiment for verifying our theoretical results.

\vspace*{12pt}

The first concern will be realistic values of $f_{\rm max}/mg$ obtained in experiments.
To estimate it, suppose that we use commercially available strings:
a sewing fiber made of polyester and a fishing line made of polyvinylidene fluoride,
both of which we are familiar with in our daily life, 
are cases in point.
These strings are typically endowed with the maximum strength, $f_{\rm max}$, of a few tens newtons (N).
For instance, the author has found a sewing thread with $f_{\rm max}=12$ N and a fishing line of $f_{\rm max}=30$ N
in a DIY store nearby.
This indicates that, if we use a weight of mass $m=500$ g in the experiment,
we have $f_{\rm max}/mg = 12/(0.5\times 9.8) \simeq 2.45$ for the sewing thread
and $f_{\rm max}/mg = 30/(0.5\times 9.8) \simeq 6.12$ for the fishing line.
These values of $f_{\rm max}/mg$ are comparable to those we have considered 
in theoretical discussion (see Figs.~\ref{fig04} and \ref{fig05}).

\vspace*{12pt}

Another concern should be the realistic value of $v^*$, which can be explored by considering 
the spring constant, $k$, of the strings used in the experiment.
Although the value of $k$ depends on the natural lengths of the strings,
we can estimate it as a few hundreds newtons per meter (N/m)
if the length of one meter or less is assumed.
In fact, the author has confirmed using a simple balance that 
the sewing thread of 50 cm in length has the spring constants of $k=170$ N/m and 
the fishing line of 1 m has $k=120$ N/m.
Therefore, if we set up the pulling speed of $V=1$ m/s and the weight of mass $m=500$ g,
we obtain
$v^* = k V/(m\omega g) = 170 \times 1/(0.5\times 26.1 \times 9.8) = 1.33$
and
$v^* = 120 \times 1/(0.5\times 21.9 \times 9.8) = 1.12$
for the sewing-thread-based apparatus and the fishing-line-based apparatus,
respectively.
What's more, since the value of $k$ can be tuned by changing the string's length,
we can cover a wide range of $v^*$ that has been considered in our theoretical discussion 

\vspace*{12pt}

The final point to be noted is how to realize a constant-pulling-speed condition.
One idea is to use a rotating wheel driven by a gear motor with a few RPM (rotations per minute).
Suppose that the bottom end of the lower string is attached to the wheel
and its rotational axis is fixed horizontally.
Then, constant rate rotation of the wheel results in a desired downward pulling with constant rate.
Another possible way is to use a freely rotating wheel made of steel, for example,
that posesses a large moment of inertia around the rotation axis.
The latter way requires to install a velocity sensor and a force sensor with the apparatus
to measure the wheel's rotation speed and the string's tension, respectively.
In either case, inexpensive and small size devices will suffice to build the experimental equipment.

\section{Conclusion}

We have theoretically described the well-known inertia demonstration based on a weight and two strings.
Under a constant-pulling-speed condition,
we have derived a closed-form expression for the critical pulling speed
at a given maximum tension.
A phase diagram illustrating which string will break has also been established.
The author thinks that it is pedagogical to follow the present argument
and compare it with those obtained under other pulling conditions \cite{LeCorbeiller1945,Karioris1978,Heald1996,Caplan2004}.

\section{Suggested problems}

It is reasonable to propose the following questions for further work related to this thesis.

\begin{itemize}

\item
Consider an upside-down situation; where you pick up the top of a string that hangs a weight,
and a similar string connects the bottom of the weight and floor.
Can you derive a closed-form expression for the critical speed of upward pulling?
What phase diagram do you obtain?

\item
What occurs if you lay down the apparatus on a friction-free floor?
In the system, the left-hand edge of the string left to the weight is fixed to a sidewall,
and the right-hand edge of the string right to the weight is what you are pulling horizontally.
Gravitational force on the weigh no longer contributes in this case.

\item
Carry out a study similar to this article for a one-dimensional chain
of many weights, in which the weights and strings are arranged alternately in the vertical direction
from above to bottom.
When pulling downward the bottom of the lowest string with a constant speed,
is it possible for all the strings to break simultaneously?
You can start a double-weight system as the simplest example.
\end{itemize}

\section*{Acknowledgements}

The author gratefully acknowledges financial support by JSPS
(the Japan Society for the Promotion of Science) KAKENHI
Grant-in-Aid for Scientific Research (C), No. 25390147.



\begin{thebibliography}{99}
\bibitem{comment}
This fact is readily proved by drawing the graphs of $y=x$ and $y=\sin(cx)$ with $0<c\le 1$.
You will find $\sin(cx)<x$ at any $x>0$.
\bibitem{sinc}
For a pedagogical merit, we comment here that
a function $(\sin x)/x$, often called by a sinc function,
is associated with spherical Bessel functions.
A spherical Bessel function of the first order is commonly denoted by
$j_{\nu}(x) = \sqrt{\pi/(2x)} J_{\nu+\frac12}(x)$
with the first-order Bessel function $J_m(x)$,
and we can prove that
$j_0(x) =  (\sin x)/x$.
\end{thebibliography}
\end{document}